\documentclass[aps,amsmath,amssymb,twocolumn,superscriptaddress,prb]{revtex4-2}
\usepackage{graphicx}
\usepackage{amsthm}
\usepackage{xcolor}
\usepackage{upquote}

\newtheorem{theorem}{Theorem}
\newtheorem{corollary}{Corollary}[theorem]
\newtheorem{lemma}[theorem]{Lemma}

\theoremstyle{definition}
\newtheorem{definition}{Definition}

\renewcommand{\r}{\vec{r}}
\renewcommand{\d}{\mathrm{d}}
\newcommand{\ve}{\varepsilon}

\usepackage{stmaryrd}
\newcommand{\lightcone}{\leftrightarrowtriangle}

\begin{document}

\title{How to detect the spacetime curvature without rulers and clocks. II. Three-dimensional spacetime}

\author{A.~V.~Nenashev}
\thanks{On leave of absence from Rzhanov Institute of Semiconductor Physics
  and the Novosibirsk State University, Russia}
\email{nenashev\_isp@mail.ru}
\affiliation{Department of Physics and Material Sciences Center,
  Philipps-Universit\"at Marburg, D-35032 Marburg, Germany}

\author {S.~D.~Baranovskii}
\email{sergei.baranovski@physik.uni-marburg.de}
\affiliation{Department of Physics and Material Sciences Center,
  Philipps-Universit\"at Marburg, D-35032 Marburg, Germany}
\affiliation{Department f\"ur Chemie, Universit\"at zu K\"oln,
  Luxemburger Stra\ss e 116,
  50939 K\"{o}ln, Germany}

\date{\today}

\begin{abstract}
	We have generalized the results of the previous work [arXiv:2302.12209] to the case of three-dimensional (3D) spacetime with two spatial and one temporal coordinates. 
	We have found that the flat Minkowski 3D spacetime is ``well-stitched'', which means that it possesses a structure described by 24 causal relations between 12 events. We have proved that a 3D spacetime is ``well-stitched'' if and only if it is conformally flat. The concept of a ``well-stitched'' spacetime does not rely on metrical information about lengths, times, etc., and does not belong to the metric geometry, but rather to geometry of incidence. We therefore have ``translated'' an important concept of a conformally-flat spacetime from the ``metric'' language of Riemannian geometry to the ``non-metric'' language of the geometry of incidence. The results of this paper provide a tool for detecting the curvature of the 3D spacetime on the basis of causal relations only, without any measurement instruments like rulers and clocks, provided that the spacetime is not conformally flat.
\end{abstract}

\maketitle   

\section{Introduction}
\label{sec:intro}

A curved spacetime is a central concept of the general theory of relativity. 
Mathematical description of the curved spacetime is given in the language of differential geometry. 
Namely, one can define an infinitesimal interval $\d s^2$ between two neighboring points as a quadratic form of the difference $\d x^i$ between coordinates of the points: $\d s^2 = g_{ik} \, \d x^i \, \d x^k$, where $g_{ik}$ is a metric tensor. Dependence of the metric tensor on coordinates defines the pseudo-Riemannian geometry of the spacetime. (The prefix ``pseudo'' reflects that there is a time coordinate in addition to spatial ones.) From the metric tensor and its derivatives one can deduce the Riemann tenor of curvature $R_{iklm}$. The Riemann tensor allows us to discriminate between a flat spacetime and a curved one: if the Riemann tensor is equal to zero in every point of the spacetime, than the spacetime is flat, otherwise it is curved.

\begin{figure}
\includegraphics[width=5cm]{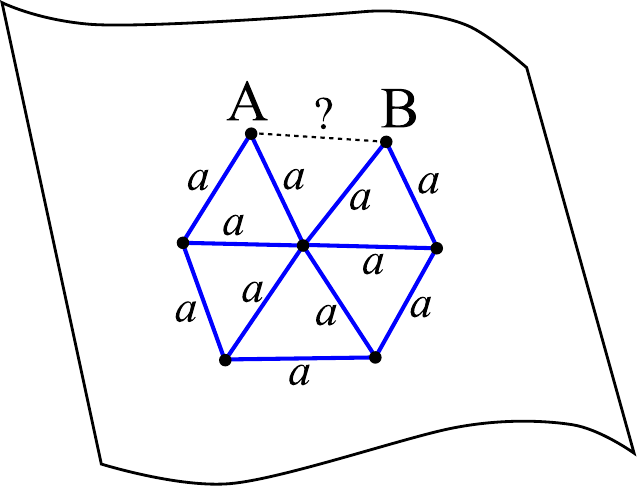}
\caption{A simple example of a finite set of measurements to detect the curvature of a surface. Blue rods, connecting points on a surface, have equal lengths $a$ on a surface. If the surface is flat, then the distance between points A and B is equal to $a$. Conversely, if the distance AB differs from $a$, then it indicates that the surface is curved.}
\label{fig:surface}
\end{figure}

This straightforward way of determining the curvature demands infinite number of measuring of intervals between infinitely-close pairs of points (i.~e., events) in the spacetime. However, one can detect the curvature with only finite number of measurements, see for example a ``five-point curvature detector'' considered by Synge~\cite[Chapter~XI, \S8]{Synge_book}. As a simple illustration of this possibility, Fig.~\ref{fig:surface} shows a construction of eleven rods of equal lengths $a$ on a two-dimensional surface. If the distance between points A and B is not equal to $a$, then the surface is curved. If, on the contrary, the distance between A and B is always equal to $a$ for every placement of this construction on the surface and every choice of length scale $a$, then the surface is flat.

In a recent paper~\cite{Nenashev2023}, we addressed the question: what can be said about curvature of a spacetime after a finite number of measurements of \emph{causal relations} between events? Is it possible to construct a scheme, analogous to that in Fig.~\ref{fig:surface}, which allows one to detect that the spacetime is curved, but without relying on any metrical information? in other words: without using any measuring instruments like rulers and clocks? This question has been answered in the case of four-dimensional (three spatial + one temporal coordinates) spacetime.

The aim of the present work is to generalize the results, obtained in Ref.~\cite{Nenashev2023}, to the case of three-dimensional spacetime, which possesses two spatial and one temporal dimensions. 


The paper is organized as follows. In Section~\ref{sec:causal}, we somewhat refine the posed question: namely, we explain that causal relations can discriminate between \emph{conformally}-flat and non-conformally flat spacetimes, rather than between flat and non-flat (curved) ones. We also introduce the Cotton tensor as a convenient tool for detecting conformally-flatness. The new results start from Section~\ref{sec:theorems}. In this section, we define the ``well-stitched spacetime'' in  three dimensions --- a concept based solely on causal relations. Then, we formulate Theorems~\ref{th:Minkowski} and~\ref{th:Cotton} that lead to the central result of this work: a 3-dimensional spacetime is conformally flat if and only if it is well-stitched. Theorems~\ref{th:Minkowski} and~\ref{th:Cotton} are proved in Sections~\ref{sec:proof1} and~\ref{sec:proof2}, respectively. Conclusions are given in Section~\ref{sec:conclusions}.

For convenience, we set the speed of light to unity. The signature adopted for the metric tensor is $(-++)$, so that in the flat 3D Minkowski spacetime, in Cartesian coordinates $t, x, y$, the metric tensor $g_{ik}$ is equal to 
\begin{equation}
\label{eq:intro:flat-metric}
\eta_{ik} =  
\begin{pmatrix}
-1 & 0 & 0 \\
 0 & 1 & 0 \\
 0 & 0 & 1
\end{pmatrix}
.
\end{equation}


\section{Causal relations and conformal maps}
\label{sec:causal}

\begin{figure}
\includegraphics[width=5cm]{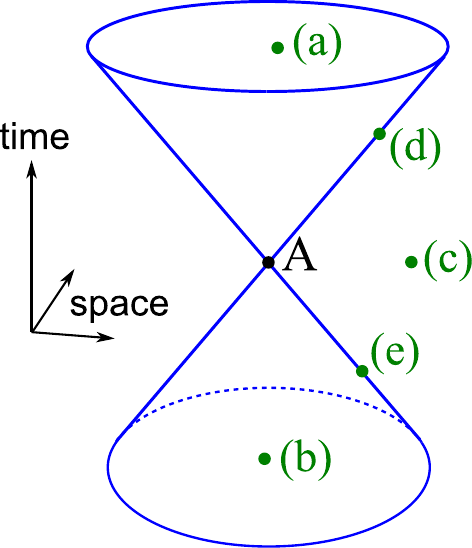}
\caption{Causal relations between two events A and B in a spacetime. Different positions of event B relative to the light cone with a vertex at A are shown by green dots. Labels (a) -- (e) correspond to different causal relations discussed in the text. In cases (d) and (e), relation ${\mathrm A} \lightcone {\mathrm B}$ is fulfilled.}
\label{fig:lightcone}
\end{figure}

A causal relation between some points (events) A and B in the spacetime may be one of the following, as illustrated in Fig.~\ref{fig:lightcone}: (a) information from A can reach B; (b) information from B can reach A; (c) information can reach neither from A to B, nor from B to A. In cases (a) and (b) the geodesic line that connects A with B is timelike, and in case (c) it is spacelike. There are also boundary cases between (a) and (c), as well as between (b) and (c), in which the geodesic line is lightlike. They are: (d) a light signal emitted from A exactly arrives at B; (e) a light signal emitted from B exactly arrives at A.

In the below text, we will combine cases (d) and (e) together, and will denote them as ${\mathrm A} \lightcone {\mathrm B}$. Relation ${\mathrm A} \lightcone {\mathrm B}$ means, in other words, that event B lies on the light cone with the vertex at event A. Equivalently, ${\mathrm A} \lightcone {\mathrm B}$ means that events A and B are connected by a lightlike geodesic line.

Examining the causal relations can provide only those information about the spacetime geometry, that is encoded in the set of light cones. It is well known that light cones are preserved under the action of a conformal map. The conformal map is a transformation of a spacetime geometry that modifies the metric $g_{ik}(\r)$ as follows:
\begin{equation}
\label{eq:intro:conformal-map}
g_{ik}(\r) \to \lambda^2(\r) \, g_{ik}(\r) ,
\end{equation}
where $\lambda(\r) \neq 0$ is an arbitrary function of coordinates. 

It is therefore impossible to distinguish, using causal relations only, between two spacetimes that are transformed one into another by a conformal map. In particular, one cannot distinguish by means of causal relations between the flat spacetime and a \emph{conformally-flat} one. A spacetime is called conformally flat if it can be mapped onto the flat Minkowski spacetime by a conformal map.

Therefore we have to refine the question posed in the title (``How to detect the spacetime curvature without rulers and clocks?'') in the following way. Whether there is such a combination of causal relations between events in a 3-dimensional spacetime, that allows us to detect the curvature \emph{if the spacetime is not conformally flat}? In Section~\ref{sec:theorems} we will present such a combination of relations, see Definition~\ref{def:1}.

To address this refined version of the question, we need a convenient criterion to recognize whether the spacetime is conformally flat. Such criterion is well known, and it depends on the dimensionality of the spacetime. In the case of four or more dimensions, one can recognize conformally-flatness by means of the Weil tensor, which is a linear combination of components of the Riemann curvature tensor $R_{iklm}$. The spacetime is conformally flat if and only if the Weil tensor is equal to zero at every point.

But in the three-dimensional spacetime the Weil tensor is always equal to zero, and therefore cannot serve as a criterion for distinguishing between conformally-flat and non-conformally flat cases. There is another tensor, which provides such a criterion in three dimensions. It is the Cotton tensor $C_{ijk}$, which is a combination of \emph{derivatives} of the curvature tensor $R_{iklm}$:
\begin{equation}
\label{eq:intro:cotton-tensor}
C_{ijk} = \partial_k R_{ij} - \partial_j R_{ik} + \frac{ \partial_j(Rg_{ik}) - \partial_k(Rg_{ij}) }{2(n-1)} \, ,
\end{equation}
where $R_{ij} = R{^l}{_{ilj}}$ is the Ricci curvature tensor, $R = R{^i}{_i}$ is the scalar curvature, and $n$ is the dimension of the spacetime. In the case of $n=3$, the spacetime is conformally flat if and only if the Cotton tensor is equal to zero at every point.

It is worth noting that \emph{two-dimensional spacetime is always conformally flat}. Therefore there is no possibility to generalize the results of this work to the two-dimensional case.


\section{Concept of a well-stitched spacetime in the 3-dimensional case}
\label{sec:theorems}

In our previous work~\cite{Nenashev2023}, we have introduced the notion of well-stitched spacetime in the four-dimensional case. Definition~\ref{def:1} below provides generalization of this notion to the case of a three-dimensional spacetime. It is based on the causal relation 
${\mathrm A} \lightcone {\mathrm B}$, which means that event B lies on the light cone with the vertex at event A.

\begin{definition}[well-stitched spacetime]
A three-dimensional spacetime is well-stitched iff for any twelve events 1, 2, 3, 4, 5, 6, $a$, $b$, $c$, $d$, $e$, and $f$, that are all different from each other, from twenty three relations 
\begin{gather*}
  a\lightcone1, \;\; a\lightcone2, \;\; a\lightcone3, \;\; a\lightcone4, \\
  b\lightcone1, \;\; b\lightcone2, \;\; b\lightcone3, \;\; b\lightcone4, \\
  c\lightcone1, \;\; c\lightcone2, \;\; c\lightcone5, \;\; c\lightcone6, \\
  d\lightcone1, \;\; d\lightcone2, \;\; d\lightcone5, \;\; d\lightcone6, \\
  e\lightcone3, \;\; e\lightcone4, \;\; e\lightcone5, \;\; e\lightcone6, \\
  f\lightcone3, \;\; f\lightcone4, \;\; f\lightcone5
\end{gather*}
follows relation $f\lightcone6$.
\label{def:1}
\end{definition}

Table \ref{table:1} clarifies this set or relations. The relation between $f$ and $6$, denoted by the question mark in the table, follows from other 23 relations denoted by symbols~$\lightcone$.
\begin{table}
\centering
\begin{tabular}{ c|cccccc| } 
     & 1 & 2 & 3 & 4 & 5 & 6 \\ 
 \hline
 $a$ & $\lightcone $ & $\lightcone $ & $\lightcone $ & $\lightcone $ & & \\ 
 $b$ & $\lightcone $ & $\lightcone $ & $\lightcone $ & $\lightcone $ & & \\ 
 $c$ & $\lightcone $ & $\lightcone $ & & & $\lightcone $ & $\lightcone $ \\ 
 $d$ & $\lightcone $ & $\lightcone $ & & & $\lightcone $ & $\lightcone $ \\ 
 $e$ & & & $\lightcone $ & $\lightcone $ & $\lightcone $ & $\lightcone $ \\ 
 $f$ & & & $\lightcone $ & $\lightcone $ & $\lightcone $ & $?$ \\ 
 \hline
\end{tabular}
\caption{Illustration of the set of relations in Definition~\ref{def:1}.}
\label{table:1}
\end{table}

\begin{figure}
\includegraphics[width=7cm]{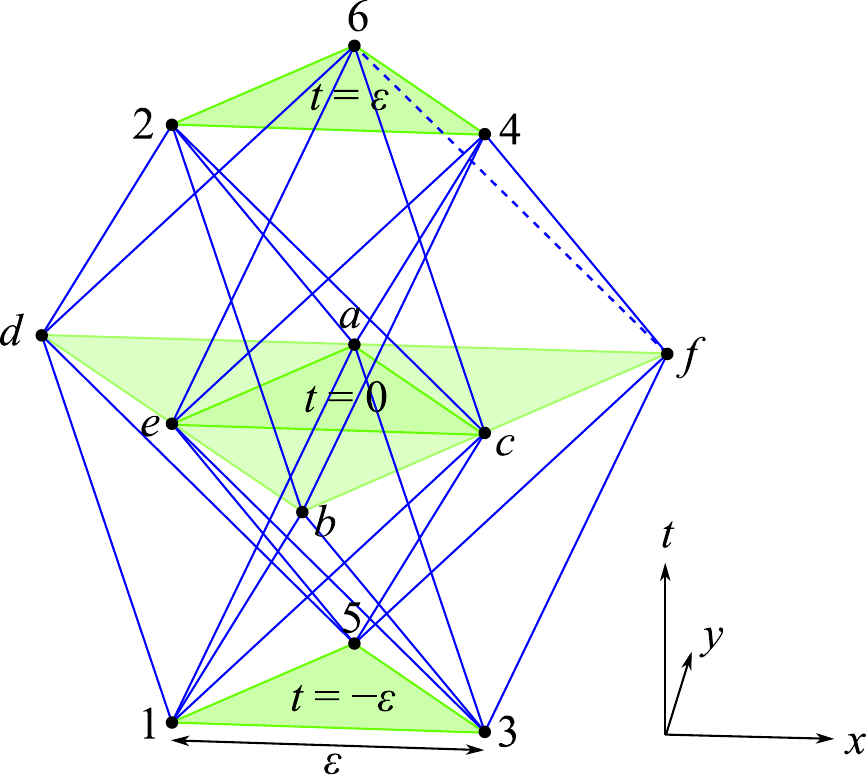}
\caption{A set of twelve events 1, 2, 3, 4, 5, 6, $a$, $b$, $c$, $d$, $e$, $f$ (black dots) in the 3D Minkowski spacetime that illustrates the relations involved in Definition~\ref{def:1} (blue lines). Coordinates $(t,x,y)$ of the events are given by Eqs.~(\ref{eq:well:135}) -- (\ref{eq:well:bdf}).}
\label{fig:simple-set}
\end{figure}

A simple example of such a set of 12 events in the 3D Minkowski spacetime is shown in Fig.~\ref{fig:simple-set}. 
It is determined by a length scale $\ve$. 
Events 1, 3 and 5 form an equilateral triangle with edge length $\ve$ in the plane $t = -\ve$. Their coordinates ($t,x,y$) are
\begin{subequations} \label{eq:well:135}
\begin{align}
&1: \; \left(-1, -\frac12, -\frac{1}{2\sqrt3}\right)\ve,  \label{eq:well:1}\\
&3: \; \left(-1, \;\;\frac12, -\frac{1}{2\sqrt3}\right)\ve,  \label{eq:well:3}\\
&5: \; \left(-1, \;\;0, \;\;\;\;\frac{1}{\sqrt3}\;\right)\ve.  \label{eq:well:5}
\end{align}
\end{subequations}
A similar triangle in plane $t = \ve$ defines events 2, 4 and~6:
\begin{subequations} \label{eq:well:246}
\begin{align}
&2: \; \left(1, -\frac12, -\frac{1}{2\sqrt3}\right)\ve, \label{eq:well:2}\\
&4: \; \left(1, \;\;\frac12, -\frac{1}{2\sqrt3}\right)\ve, \label{eq:well:4}\\
&6: \; \left(1, \;\;0, \;\;\;\;\frac{1}{\sqrt3}\;\right)\ve. \label{eq:well:6}
\end{align}
\end{subequations}
Events $e$, $c$ and $a$ also form a similar triangle that lies in plane $t=0$, exactly between triangles 135 and 246:
\begin{subequations} \label{eq:well:eca}
\begin{align}
&e: \; \left(0, -\frac12, -\frac{1}{2\sqrt3}\right)\ve, \label{eq:well:e}\\
&c: \; \left(0, \;\;\frac12, -\frac{1}{2\sqrt3}\right)\ve, \label{eq:well:c}\\
&a: \; \left(0, \;\;0, \;\;\;\;\frac{1}{\sqrt3}\;\right)\ve, \label{eq:well:a}
\end{align}
\end{subequations}
and the rest three events $b$, $d$ and $f$ also lie in plane $t=0$ and form a larger equilateral triangle, in which the midpoints of edges coincide with events $e$, $c$ and $a$:
\begin{subequations} \label{eq:well:bdf}
\begin{align}
&b: \; \left(0, \;\;0, -\frac{2}{\sqrt3}\right)\ve, \label{eq:well:b}\\
&d: \; \left(0, -1,    \;\;\frac{1}{\sqrt3}\right)\ve, \label{eq:well:d}\\
&f: \; \left(0, \;\;1, \;\;\frac{1}{\sqrt3}\right)\ve. \label{eq:well:f}
\end{align}
\end{subequations}
One can easily see that for every pair of events marked by sign $\lightcone$ in Table~\ref{table:1}, for example $a$ and 1, the spatial difference is equal to $\ve$, and the difference in time is also $\ve$. Since we imply that the speed of light is unity, then a light signal, emitted at the earlier event, arrives exactly at the later event, that is the relation $\lightcone$ is fulfilled between these events, e.~g. $a\lightcone1$. In Fig.~\ref{fig:simple-set}, trajectories of light signals in such pairs are shown by blue solid lines. Relation $f\lightcone6$ (marked as ``?'' in Table~\ref{table:1} and as a dashed line in Fig.~\ref{fig:simple-set}) is also fulfilled, suggesting that the Minkowski spacetime may be well-stitched.
And indeed, the following theorem holds:

\begin{theorem}
The 3-dimensional Minkowski spacetime is well-stitched.
\label{th:Minkowski}
\end{theorem}

A proof of Theorem~\ref{th:Minkowski} is given in Section~\ref{sec:proof1}. Within this proof, we will see that the combination of events given by Eqs.~(\ref{eq:well:135}) -- (\ref{eq:well:bdf}) is not unique. There are many other arrangements of events that obey relations $a\lightcone1, \ldots, f\lightcone5$ listed in Definition~\ref{def:1} as well as in Table~\ref{table:1}. As we show in Section~\ref{sec:proof1}, relation $f\lightcone6$ turns out to be fulfilled for each of these arrangements. This observation consists the proof of Theorem~\ref{th:Minkowski}.

As we discussed in Section~\ref{sec:causal}, causal relations do not distinguish between a flat and a conformally-flat spacetime. Since the concept of well-stitched spacetime is based solely on causal relation, one can generalize the statement of Theorem~\ref{th:Minkowski} to the case of a conformally-flat spacetime as follows:

\begin{corollary}
Any 3-dimensional conformally flat spacetime is well-stitched.
\label{cor:Minkowski}
\end{corollary}

Now we know from Corollary~\ref{cor:Minkowski} how the statement that a spacetime is conformally flat can be ``translated'' from the language of differential geometry to the language of causal relations. For the sake of completeness, we would like to have analogous ``translation'' also in the situation when a spacetime is not conformally flat. Such a possibility is provided by the following Theorem, which will be proved in Section~\ref{sec:proof2}:

\begin{theorem}
If a 3-dimensional spacetime has a non-zero Cotton tensor at some point $O$, then well-stitchedness violates in some vicinity of $O$.
\label{th:Cotton}
\end{theorem}

As we mentioned in Section~\ref{sec:causal}, the three-dimensional spacetime is conformally flat if and only if the Cotton tensor is equal to zero at every point. Hence, if the spacetime is not conformally flat, then the Cotton tensor must differ from zero at some point. Applying Theorem~\ref{th:Cotton}, we arrive at the following result:

\begin{corollary}
If a 3-dimensional spacetime is not conformally flat, then it is not well-stitched.
\label{cor:Cotton}
\end{corollary}

Corollaries \ref{cor:Minkowski} and \ref{cor:Cotton} together consist the main result of this paper:\newline
\textbf{A 3-dimensional spacetime is conformally flat if and only if it is well-stitched.}


\section{Causal structure of the flat spacetime: proof of Theorem~\ref{th:Minkowski}}
\label{sec:proof1}

In this section, we consider events in a flat 3-dimensional Minkowski space. The interval $ds^2_{PQ}$ between two events $P = (t_P,x_P,y_P)$ and $Q = (t_Q,x_Q,y_Q)$ is defined as
\begin{equation}
\label{eq:1:interval}
ds_{PQ}^2 = -(t_P-t_Q)^2 + (x_P-x_Q)^2 + (y_P-y_Q)^2,
\end{equation}
and relation $P\lightcone Q$ just means that $ds_{PQ}^2 = 0$. 

We also define the scalar product of two vectors $\vec{v} = (v_t,v_x,v_y)$ and $\vec{w} = (w_t,w_x,w_y)$ as
\begin{equation}
\label{eq:1:scalar_product}
\vec{v} \cdot \vec{w} = -v_tw_t + v_xw_x + v_yw_y \, ,
\end{equation}
and the square of a vector $\vec{v} = (v_t,v_x,v_y)$ as
\begin{equation}
\label{eq:1:square}
(\vec{v})^2 = \vec{v} \cdot \vec{v} = -v_t^2 + v_x^2 + v_y^2 \, .
\end{equation}

We imply in this section that twelve points (events) 1, 2, 3, 4, 5, 6, $a$, $b$, $c$, $d$, $e$, $f$ are all different from each other, and that 23 relations $a\lightcone1, \ldots, f\lightcone5$ listed in Definition~\ref{def:1} are fulfilled. The aim of this section is to prove that relation $f\lightcone6$ is also fulfilled under these conditions.

\begin{lemma}
\label{lemma:one-plane}
In a flat 3-dimensional Minkowski space, under conditions of Definition~\ref{def:1}, points {\rm1}, {\rm2}, {\rm3}, and {\rm4} lie in one plane.
\end{lemma}

{\it Proof of Lemma~\ref{lemma:one-plane}.} The interval $ds_{ab}^2$ between points $a$ and $b$ can be either timelike ($ds_{ab}^2<0$), or spacelike ($ds_{ab}^2>0$), or lightlike ($ds_{ab}^2=0$). We consider these three options separately.

\begin{figure}
\includegraphics[width=\linewidth]{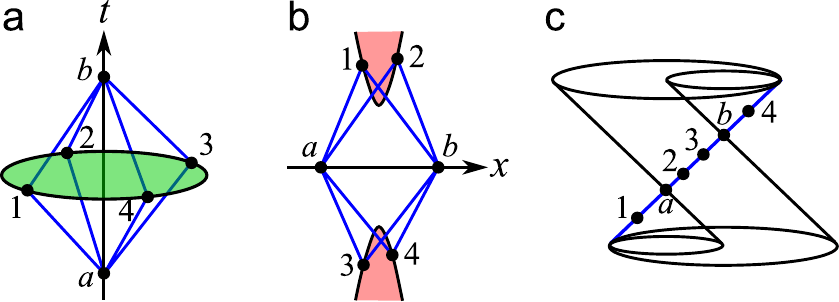}
\caption{Illustration to the proof of Lemma~\ref{lemma:one-plane}: (a) case of $ds_{ab}^2<0$, (b) case of $ds_{ab}^2>0$, (c) case of $ds_{ab}^2=0$.}
\label{fig:lemma}
\end{figure}

If $ds_{ab}^2<0$, then one can choose line $ab$ as the axis of time $t$, and put the origin of coordinates to the midpoint of the line segment $ab$, so that $t_a = -t_b$. It follows from relations $a\lightcone1$ and $b\lightcone1$ that point 1 lies at the intersection of two light cones with vertices at $a$ and at $b$. This intersection is a circle lying in plane $t=0$, as illustrated in Fig.~\ref{fig:lemma}a. Also, due to relations $a\lightcone2$, $b\lightcone2$, $a\lightcone3$, $b\lightcone3$, $a\lightcone4$, and $b\lightcone4$, points 2, 3, and 4 lie at the same circle. Therefore, all four points 1, 2, 3, and 4 belong to the same plane $t=0$.

If $ds_{ab}^2<0$, then one can choose line $ab$ as the spatial $x$-axis, and put the origin of coordinates to the midpoint of the line segment $ab$, so that $x_a = -x_b$. Just as in the previous case, four points 1, 2, 3, 4 lie at the intersection of two light cones: one with the vertex at $a$, and another with the vertex at $b$ (Fig.~\ref{fig:lemma}b). This intersection is a hyperbola lying in plane $x=0$. Hence, points 1, 2, 3, and 4 belong to plane $x=0$.

Finally, if $ds_{ab}^2=0$, then two light cones with vertices at $a$ and at $b$ intersect at the line $ab$ (Fig.~\ref{fig:lemma}c). Consequently, points 1, 2, 3 and 4 must lie on this line.

Summarizing, we see that in each of three cases the points 1, 2, 3, and 4 lie in one plane.\hfill$\blacksquare$

Now we return to the proof of Theorem~\ref{th:Minkowski}. Consider two lines 12 and 34, the first joins points 1 and 2, and the second joins points 3 and 4. Line 12 can be either timelike, or spacelike, or lightlike. The same is true for line 34. Due to Lemma~\ref{lemma:one-plane}, lines 12 and 34 lie in the same plane, so that they can either intersect at some point, or be parallel to each other, or coincide. We divide the set of all these possibilities into four cases:
\begin{itemize}
\item Case A: at least one of two lines 12 and 34 is lightlike.
\item Case B: lines 12 and 34 are parallel to each other or coincide, and both are timelike.
\item Case C: lines 12 and 34 are parallel to each other or coincide, and both are spacelike.
\item Case D: lines 12 and 34 intersect at some point, and none of them is lightlike. (This is the generic case.)
\end{itemize}

Let us show that relation $f\lightcone6$ is fulfilled in each of these cases.

{\it Case A.} Suppose that line 12 is lightlike. It follows from relations $a\lightcone1$, $a\lightcone2$, $b\lightcone1$, $b\lightcone2$, $c\lightcone1$, $c\lightcone2$, $d\lightcone1$, $d\lightcone2$ that points $a$, $b$, $c$ and $d$ lie at the intersection of two light cones with vertices at points $1$ and $2$. This intersection is line 12 (cf. Fig.~\ref{fig:lemma}c). Therefore, line 12 contains not only points 1 and 2, but also points $a$, $b$, $c$ and $d$. This argument can be continued further: relations $a\lightcone3$, $a\lightcone4$, $b\lightcone3$, $b\lightcone4$ mean that points 3 and 4 lie at the intersection of light cones with vertices at $a$ and at $b$, that is, on the same line 12. Then, it follows from relations $c\lightcone5$, $c\lightcone6$, $d\lightcone5$, $d\lightcone6$ that points 5 and 6 also belong to line 12. Finally, it follows from relations $e\lightcone3$, $e\lightcone4$, $f\lightcone3$ and $f\lightcone4$ that points $e$ and $f$ lie on the same line. One can therefore conclude that all twelve points 1, 2, 3, 4, 5, 6, $a$, $b$, $c$, $d$, $e$ and $f$ belong to the same light-like line. Relation $f\lightcone6$ follows then from the fact that points $f$ and 6 are connected by a light-like line.

The same reasoning is applied if we suppose that line 34 is lightlike. Namely, we use relations $a\lightcone3$, $a\lightcone4$, $b\lightcone3$, $b\lightcone4$, $e\lightcone3$, $e\lightcone4$, $f\lightcone3$ and $f\lightcone4$ to ensure that points $a$, $b$, $e$ and $f$ belong to line 34. Then, relations $a\lightcone1$, $a\lightcone2$, $b\lightcone1$, $b\lightcone2$ prove that points 1 and 2 lie on the same line. It follows then from relations $c\lightcone1$, $c\lightcone2$, $d\lightcone1$ and $d\lightcone2$ that points $c$ and $d$ also lie on this line. And finally, one can conclude from relations $c\lightcone5$, $c\lightcone6$, $d\lightcone5$ and $d\lightcone6$ that points 5 and 6 belong to the same line. Therefore all the twelve points lie on one lightlike line, that entails relation $f\lightcone6$.

\begin{figure}
\includegraphics[width=\linewidth]{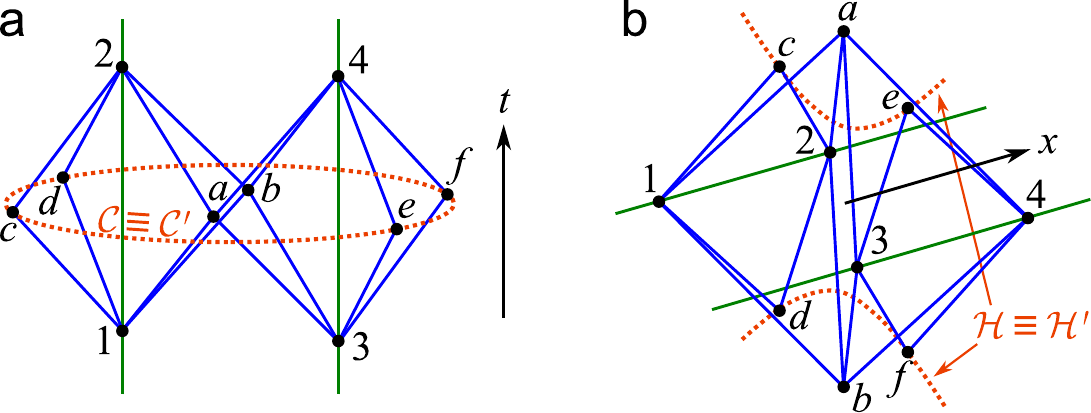}
\caption{Illustration to the proof of Theorem~\ref{th:Minkowski}: (a) case B of timelike parallel lines 12 and 34, (b) case C of spacelike parallel lines 12 and 34. Lines 12 and 34 are displayed in green. Lightlike intervals between events are shown as blue lines. Circle $\mathcal C$, which coincides with circle $\mathcal C'$, is depicted in orange, as well as hyperbola $\mathcal H$, which is the same as hyperbola $\mathcal H'$.}
\label{fig:casesBC}
\end{figure}

{\it Case B.} Since lines 12 and 34 are timelike and parallel to each other, we can choose the axis of time $t$ along these lines. Due to relations $a\lightcone1$, $a\lightcone2$, $b\lightcone1$, $b\lightcone2$, $c\lightcone1$, $c\lightcone2$, $d\lightcone1$ and $d\lightcone2$, four points $a$, $b$, $c$, $d$ are located on the intersection of two light cones with vertices at point 1 and at point 2. This intersection belongs to the $xy$-plane that lies in the middle between points 1 and 2 (see Fig.~\ref{fig:casesBC}a). Therefore $t$-coordinates of points $a$, $b$, $c$ and $d$ are the same:
\begin{equation}
\label{eq:1:caseB-tabcd}
t_a = t_b = t_c = t_d \, .
\end{equation}
Similarly, it follows from relations $a\lightcone3$, $a\lightcone4$, $b\lightcone3$, $b\lightcone4$, $e\lightcone3$, $e\lightcone4$, $f\lightcone3$ and $f\lightcone4$ that
\begin{equation}
\label{eq:1:caseB-tabef}
t_a = t_b = t_e = t_f \, .
\end{equation}
Hence, six points $a$, $b$, $c$, $d$, $e$ and $f$ lie in the same $xy$-plane.

Then, let us consider relations $c\lightcone5$, $d\lightcone5$, $e\lightcone5$ and $f\lightcone5$. They mean that points $c$, $d$, $e$ and $f$ lie on the light cone with the vertex at point 5. At the same time, points $c$, $d$, $e$ and $f$ lie in one $xy$-plane. Therefore, points $c$, $d$, $e$ and $f$ belong to some circle $\mathcal C$ --- an intersection of the light cone and the plane. Similarly, one can conclude on the basis of relations $c\lightcone6$, $d\lightcone6$ and $e\lightcone6$ that points $c$, $d$ and $e$ lie on some circle $\mathcal C'$. Moreover, for any point $p$ lying on circle $\mathcal C'$, relation $p\lightcone6$ fulfills since the whole circle $\mathcal C'$ belongs to the light cone with the vertex at point 6. Circle $\mathcal C$ is the same as circle $\mathcal C'$, because each of them contains the same triple of different points $c$, $d$ and $e$. Hence, point $f$ lies on circle $\mathcal C'$, and consequently the relation $f\lightcone6$ fulfills.

{\it Case C.} Consideration of this case is essentially the same as that of Case B, with minor variations. Since lines 12 and 34 are spacelike and parallel to each other, we can choose the spatial $x$-axis along these lines. Due to relations $a\lightcone1$, $a\lightcone2$, $b\lightcone1$, $b\lightcone2$, $c\lightcone1$, $c\lightcone2$, $d\lightcone1$ and $d\lightcone2$, four points $a$, $b$, $c$, $d$ are located on the intersection of two light cones with vertices at point 1 and at point 2. This intersection belongs to the $ty$-plane that lies in the middle between points 1 and 2 (see Fig.~\ref{fig:casesBC}b). Therefore $x$-coordinates of points $a$, $b$, $c$ and $d$ are the same:
\begin{equation}
\label{eq:1:caseC-xabcd}
x_a = x_b = x_c = x_d \, .
\end{equation}
Similarly, it follows from relations $a\lightcone3$, $a\lightcone4$, $b\lightcone3$, $b\lightcone4$, $e\lightcone3$, $e\lightcone4$, $f\lightcone3$ and $f\lightcone4$ that
\begin{equation}
\label{eq:1:caseC-xabef}
x_a = x_b = x_e = x_f \, .
\end{equation}
Hence, six points $a$, $b$, $c$, $d$, $e$ and $f$ lie in the same $ty$-plane.

Then, let us consider relations $c\lightcone5$, $d\lightcone5$, $e\lightcone5$ and $f\lightcone5$. They mean that points $c$, $d$, $e$ and $f$ lie on the light cone with the vertex at point 5. At the same time, points $c$, $d$, $e$ and $f$ lie in one $ty$-plane. Therefore, points $c$, $d$, $e$ and $f$ belong to some hyperbola $\mathcal H$ --- an intersection of the light cone and the plane. This hyperbola is described by equation
\begin{equation}
\label{eq:1:caseC-hyperbola}
-(t-t_0)^2 + (y-y_0)^2 + r^2 = 0,  \quad x = \text{const} ,
\end{equation}
where $t_0$, $y_0$ and $r$ are some constants. 
Similarly, one can conclude on the basis of relations $c\lightcone6$, $d\lightcone6$ and $e\lightcone6$ that points $c$, $d$ and $e$ lie on some hyperbola $\mathcal H'$, which also has a shape described by Eq.~(\ref{eq:1:caseC-hyperbola}). For any point $p$ lying on hyperbola $\mathcal H'$, relation $p\lightcone6$ fulfills. Then, one can deduce that hyperbolas $\mathcal H$ and $\mathcal H'$ are the same from the fact that each of them contains the same triple of different points $c$, $d$ and $e$. Hence, point $f$ lies on hyperbola $\mathcal H'$, and consequently the relation $f\lightcone6$ fulfills.

{\it Case D.} Let us denote the point of intersection of lines 12 and 34 as $X$. We choose $X$ as an origin of coordinates. For any point $p$, its radius-vector $\vec{r}_p$ is therefore a vector that joins $X$ with $p$. 

It is easy to show that none of points 1, 2, 3, 4 coincides with $X$. Indeed, if for example point 1 is the same as $X$, then line 34 contains points 1, 3 and 4. According to relations $a\lightcone1$, $a\lightcone3$ and $a\lightcone4$, three different points 1, 3 and 4 lie on the light cone with the vertex at $a$, and simultaneously these three points lie on line 34. This means that line 34 entirely belongs to this light cone, and therefore is lightlike. But this contradicts to the premise of Case D that line 34 is not lightlike. Hence, point 1 is different from point $X$. The same reasoning is valid for points 2, 3 and 4. As a result, radius-vectors $\vec{r}_1$, $\vec{r}_2$, $\vec{r}_3$, $\vec{r}_4$ of points 1, 2, 3, 4 are all nonzero.

Since vectors $\vec{r}_1$ and $\vec{r}_2$ are collinear and both nonzero, they are proportional to each other:
\begin{equation}
\label{eq:1:caseD-lambda}
\vec{r}_2 = \lambda \vec{r}_1 \, ,
\end{equation}
where $\lambda \neq 0$ because $\vec{r}_2 \neq 0$, and $\lambda \neq 1$ because points 1 and 2 are different. 
Relations $a\lightcone1$ and $a\lightcone2$ in a flat spacetime mean that the corresponding intervals are equal to zero: $(\vec{r}_1-\vec{r}_a)^2=0$ and $(\vec{r}_2-\vec{r}_a)^2=0$. Expanding the brackets and substituting $\lambda \vec{r}_1$ instead of $\vec{r}_2$, one obtains:
\begin{equation}
\label{eq:1:caseD-interval-a1}
(\vec{r}_1)^2 - 2\vec{r}_1\cdot\vec{r}_a + (\vec{r}_a)^2 = 0 ,
\end{equation}
\begin{equation}
\label{eq:1:caseD-interval-a2}
\lambda^2(\vec{r}_1)^2 - 2\lambda\vec{r}_1\cdot\vec{r}_a + (\vec{r}_a)^2 = 0 .
\end{equation}
One can exclude scalar product $\vec{r}_1\cdot\vec{r}_a$ multiplying Eq.~(\ref{eq:1:caseD-interval-a1}) by $\lambda$ and subtracting Eq.~(\ref{eq:1:caseD-interval-a2}):
\begin{equation}
\label{eq:1:caseD-interval-ex-scalar-product}
(\lambda-\lambda^2)(\vec{r}_1)^2  + (\lambda-1)(\vec{r}_a)^2 = 0 .
\end{equation}
Then, contracting the factor $(\lambda-1)$, one finds that
\begin{equation}
\label{eq:1:caseD-ra2}
(\vec{r}_a)^2 = \lambda (\vec{r}_1)^2 .
\end{equation}
The right-hand side of this equation is different from zero since $\lambda \neq 0$ and $(\vec{r}_1)^2 \neq 0$ (the latter inequality arises from the fact that $\vec{r}_1$ is a nonzero vector directed along non-lightlike line 12). Hence, 
\begin{equation}
\label{eq:1:caseD-ra2-nonzero}
(\vec{r}_a)^2 \neq 0 .
\end{equation}

Similarly, relations $b\lightcone1$, $b\lightcone2$, $c\lightcone1$, $c\lightcone2$, $d\lightcone1$ and $d\lightcone2$ give rise to equalities
\begin{equation}
\label{eq:1:caseD-rbcd2}
(\vec{r}_b)^2 = (\vec{r}_c)^2 = (\vec{r}_d)^2 = \lambda (\vec{r}_1)^2.
\end{equation}
These considerations can also be applied to relations $a\lightcone3$, $a\lightcone4$, $b\lightcone3$, $b\lightcone4$, $e\lightcone3$, $e\lightcone4$, $f\lightcone3$ and $f\lightcone4$, just by replacing points 1, 2, $a$, $b$, $c$, $d$ with points 3, 4, $a$, $b$, $e$, $f$. As a result,
\begin{equation}
\label{eq:1:caseD-rabef2}
(\vec{r}_a)^2 = (\vec{r}_b)^2 = (\vec{r}_e)^2 = (\vec{r}_f)^2.
\end{equation}

Summarizing Eqs.~(\ref{eq:1:caseD-ra2}) -- (\ref{eq:1:caseD-rabef2}), one can see that
\begin{equation}
\label{eq:1:caseD-rabcdef2}
(\vec{r}_a)^2 = (\vec{r}_b)^2 = (\vec{r}_c)^2 = (\vec{r}_d)^2 = (\vec{r}_e)^2 = (\vec{r}_f)^2 \neq 0.
\end{equation}

Let us prove now that points $c$, $d$ and $e$ do not lie on one line. According to Eq.~(\ref{eq:1:caseD-rabcdef2}), these points lie on some hyperboloid described by equation
\begin{equation}
\label{eq:1:caseD-hyperboloid}
-t^2 + x^2 + y^2 = \xi ,
\end{equation}
where $\xi\neq0$. If it is a two-sheet hyperboloid ($\xi<0$), then it cannot intersect with a line in more than two points, and hence three different points $c$, $d$ and $e$ on the hyperboloid cannot lie on one line. If it is a one-sheet hyperboloid ($\xi>0$), and some line has three common points $c$, $d$, $e$ with it, then this line lies entirely on the hyperboloid. Any such line is lightlike. That is, line $cd$ that joins points $c$ and $d$ is lightlike. Then, one can see from relations $c\lightcone1$, $c\lightcone2$, $d\lightcone1$ and $d\lightcone2$ that points 1 and 2 belong to the intersection of two light cones with vertices at $c$ and at $d$. In the case of lightlike line $cd$, this intersection is line $cd$. Hence, line 12 that joins points 1 and 2 is the same as the lightlike line $cd$. But it contradicts to the premise of Case D that line 12 is not lightlike. We therefore see that points $c$, $d$ and $e$ do not lie on one line.

Now consider relation $c\lightcone5$. It means that the interval between points $c$ and 5 is equal to zero: 
\begin{equation}
\label{eq:1:caseD-interval-c5}
(\vec{r}_5-\vec{r}_c)^2=0 .
\end{equation}
First, we note that, due to this relation, vector $\vec{r}_5$ differs from zero. Indeed, if $\vec{r}_5$ were equal to zero, then Eq.~(\ref{eq:1:caseD-interval-c5}) would take the form $(\vec{r}_c)^2=0$ that contradicts to Eq.~(\ref{eq:1:caseD-rabcdef2}). Second, expanding the brackets in Eq.~(\ref{eq:1:caseD-interval-c5}), one can express scalar product $\vec{r}_5\cdot\vec{r}_c$ as
\begin{equation}
\label{eq:1:caseD-r5rc}
\vec{r}_5\cdot\vec{r}_c = \frac{ (\vec{r}_5)^2 + (\vec{r}_c)^2 }{2} \, .
\end{equation}
Similarly, one can obtain from relations $d\lightcone5$, $e\lightcone5$ and $f\lightcone5$ that
\begin{equation}
\label{eq:1:caseD-r5rd}
\vec{r}_5\cdot\vec{r}_d = \frac{ (\vec{r}_5)^2 + (\vec{r}_d)^2 }{2} \, ,
\end{equation}
\begin{equation}
\label{eq:1:caseD-r5re}
\vec{r}_5\cdot\vec{r}_e = \frac{ (\vec{r}_5)^2 + (\vec{r}_e)^2 }{2} \, ,
\end{equation}
\begin{equation}
\label{eq:1:caseD-r5rf}
\vec{r}_5\cdot\vec{r}_f = \frac{ (\vec{r}_5)^2 + (\vec{r}_f)^2 }{2} \, .
\end{equation}
Right-hand parts of equations~(\ref{eq:1:caseD-r5rc}) -- (\ref{eq:1:caseD-r5rf}) are equal to each other due to Eq.~(\ref{eq:1:caseD-rabcdef2}). Hence,
\begin{equation}
\label{eq:1:caseD-r5rcdef}
\vec{r}_5\cdot\vec{r}_c = \vec{r}_5\cdot\vec{r}_d = \vec{r}_5\cdot\vec{r}_e = \vec{r}_5\cdot\vec{r}_f \, .
\end{equation}

Similarly, relations $c\lightcone6$, $d\lightcone6$ and $e\lightcone6$ entail that vector $\vec{r}_6$ is nonzero and that
\begin{equation}
\label{eq:1:caseD-r6rcde}
\vec{r}_6\cdot\vec{r}_c = \vec{r}_6\cdot\vec{r}_d = \vec{r}_6\cdot\vec{r}_e  \, .
\end{equation}

Due to Eq.~(\ref{eq:1:caseD-r5rcdef}), vector $\vec{r}_5$ is orthogonal to two linearly-independent vectors $\vec{r}_d-\vec{r}_c$ and $\vec{r}_e-\vec{r}_c$ of plane $cde$:
\begin{equation}
\label{eq:1:caseD-r5-vs-cde}
\vec{r}_5\cdot(\vec{r}_d-\vec{r}_c) = \vec{r}_5\cdot(\vec{r}_e-\vec{r}_c) = 0 .
\end{equation}
This means that vector $\vec{r}_5$ is perpendicular to plane $cde$. Analogously, it follows from Eq.~(\ref{eq:1:caseD-r6rcde}) that vector $\vec{r}_6$ is perpendicular to the same plane $cde$. Therefore vectors $\vec{r}_5$ and $\vec{r}_6$ are collinear. Since both of them are nonzero, there is a number $\mu$ such that
\begin{equation}
\label{eq:1:caseD-r5-vs-r6}
\vec{r}_6 = \mu \vec{r}_5 \, .
\end{equation}
Multiplying the last equality in Eq.~(\ref{eq:1:caseD-r5rcdef}) by $\mu$ and using Eq.~(\ref{eq:1:caseD-r5-vs-r6}), one can see that
\begin{equation}
\label{eq:1:caseD-r6ref}
\vec{r}_6\cdot\vec{r}_e = \vec{r}_6\cdot\vec{r}_f \, .
\end{equation}
Finally, from Eqs.~(\ref{eq:1:caseD-rabcdef2}) and (\ref{eq:1:caseD-r6ref}) it follows that
\begin{multline}
\label{eq:1:caseD-intervals-r6e-r6f}
(\vec{r}_6-\vec{r}_e)^2 = (\vec{r}_6)^2 - 2\vec{r}_6\cdot\vec{r}_e + (\vec{r}_e)^2 \\
= (\vec{r}_6)^2 - 2\vec{r}_6\cdot\vec{r}_f + (\vec{r}_f)^2 = (\vec{r}_6-\vec{r}_f)^2 .
\end{multline}
The left-hand side of Eq.~(\ref{eq:1:caseD-intervals-r6e-r6f}) is equal to zero due to relation $e\lightcone6$. Consequently, $(\vec{r}_6-\vec{r}_f)^2=0$, that is, relation $f\lightcone6$ is fulfilled.

Now we have considered all cases A, B, C and D. In any case, relation $f\lightcone6$ always follows from other twenty three relations $a\lightcone1, \ldots, f\lightcone5$ between twelve different events 1, 2, 3, 4, 5, 6, $a$, $b$, $c$, $d$, $e$, $f$ in a 3-dimensional Minkowski spacetime. Theorem~\ref{th:Minkowski} is therefore proved.


\section{Causal structure of a spacetime with a nonzero Cotton tensor: proof of Theorem~\ref{th:Cotton}}
\label{sec:proof2}

\subsection{Cotton-York tensor}
\label{sec:york}

For our goal, it is convenient to convert the Cotton tensor $C_{ijk}$ into its dual form, the so-called Cotton-York tensor $Y{^i}{_j}$ defined as
\begin{equation}
\label{eq:2:york-def}
Y{^i}{_j} = -\frac12 \, \varepsilon^{ikl} C_{jkl} \, ,
\end{equation}
where $\varepsilon^{ikl}$ is the Levi-Civita symbol. (Strictly speaking, $Y{^i}{_j}$ is a tensor density.) If the Cotton tensor $C_{ijk}$ differs from zero at some point $O$, so does the Cotton-York tensor $Y{^i}{_j}$.

Tensor $Y_{ij}$ is known to be symmetric ($Y_{ij} = Y_{ji}$) and traceless ($Y{^i}{_i} = 0$). A tensor possessing these two properties realizes a five-dimensional irreducible representation of the Lorentz group. All five degrees of freedom of this tensor turn into each other under Lorentz transformations. In particular, if tensor $Y_{ij}$ is non-zero as a whole, but $Y_{tt} = 0$, then it is always possible to make the component $Y_{tt}$ non-zero by applying an appropriate Lorentz transformation. Indeed, if $Y_{tt}$ were equal to zero after any Lorentz transformation, then other degrees of freedom would consist a representation of the Lorentz group with a dimension less than five, which would contradict to irreducibility of the representation given by tensor $Y_{ij}$. We will use this finding in the next subsection.

\subsection{Choice of a reference frame}
\label{sec:frame}

According to the premise of Theorem~\ref{th:Cotton}, we suppose that the Cotton tensor differs from zero at some event $O$. 

In the rest of this section, we choose so-called \emph{Riemann normal coordinates}~\cite[\S11.6]{Wheeler_book} with the origin at event $O$. In these coordinates, metric tensor $g_{ik}$ at the origin is the same as in Cartesian coordinates of a flat spacetime, and the first derivatives of $g_{ik}$ vanish at the same event:
\begin{gather}
\label{eq:2:g-is-eta}
g_{ik} = \eta_{ik} \text{ at } O, \\
\label{eq:2:g-deriv-zero}
\frac{\partial g_{ik}}{\partial r^l} = 0 \text{ at } O, 
\end{gather}
where $\eta_{ik}$ is defined by Eq.~(\ref{eq:intro:flat-metric}).
Second and third derivatives of $g_{ik}$ at event $O$ are determined by the curvature tensor $R_{iklm}$ and its derivatives, such that~\cite{Brewin2009}
\begin{multline}
\label{eq:2:g-series}
g_{ik}(\r) = \eta_{ik} - \frac13 R_{ilkm}(O) \, r^l r^m \\
- \frac16 \left[\partial_n R_{ilkm}(O)\right] r^l r^m r^n 
+ \mathcal O\left(|\r|^4\right),
\end{multline}
where $|\r| = \sqrt{t^2+x^2+y^2}$.

Additionally, we demand that
\begin{equation}
\label{eq:2:C-component-nonzero}
Y_{tt}(O) \neq 0.
\end{equation}
In section~\ref{sec:york}, we have found out that this inequality always can be fulfilled by performing some Lorentz transformation, provided that the Cotton tensor differs from zero at event $O$.

In coordinates specified by Eq.~(\ref{eq:2:g-series}), the component $Y_{tt}$ of the Cotton-York tensor at event $O$ can be easily calculated on the basis of Eqs.~(\ref{eq:intro:flat-metric}), (\ref{eq:intro:cotton-tensor}), (\ref{eq:2:york-def}), (\ref{eq:2:g-is-eta}), and (\ref{eq:2:g-deriv-zero}):
\begin{multline}
\label{eq:2:Ytt-at-O}
Y_{tt}(O) = \partial_y R_{xt}(O) - \partial_x R_{yt}(O) \\ 
= \partial_y R_{yxyt}(O) - \partial_x R_{xyxt}(O) .
\end{multline}

\subsection{Intervals in a neighborhood of event $O$}
\label{sec:intervals}

To decide whether relation $P\lightcone Q$ is fulfilled in a curved spacetime, we cannot rely on the special-relativity expression $ds_{PQ}^2 = -(t_P-t_Q)^2 + (x_P-x_Q)^2 + (y_P-y_Q)^2$ for the interval $ds_{PQ}^2$. Instead of this, we can define an ``interval'' $\delta s_{PQ}^2$ by integration of the line element $(g_{ik} dr^i dr^k)^{1/2}$ along the geodesic line that connects two given events:
\begin{equation}
\label{eq:2:delta-s-def}
\delta s^2_{PQ} = \pm \left[ \int_P^Q \sqrt{\pm g_{ik} \frac{dr^i}{d\lambda} \, \frac{dr^k}{d\lambda}  } \;  d\lambda  \right]^2 ,
\end{equation}
where integration is performed over the geodesic line segment between $P$ and $Q$, $\lambda$ is a parameter along this geodesic line, $r^i(\lambda)$ are coordinates of a point on the geodesic line, and two signs $\pm$ are equal to each other and chosen such that the expression under the square root is non-negative. 

This quantity $\delta s^2_{PQ}$ is sometimes referred as the square of the geodesic arc length. Also $\delta s^2_{PQ}$ is Synge's world function $\Omega(PQ)$~\cite{Synge_book} multiplied by 2. If the geodesic line that connects $P$ and $Q$ is timelike, than $\delta s^2_{PQ} = -(\tau_{PQ})^2$, where $\tau_{PQ}$ is the proper time passed between events $P$ and $Q$ along the geodesic. If this geodesic line is spacelike, than $\delta s^2_{PQ}$ is the squared length of the geodesic line segment between $P$ and $Q$. And if this geodesic line is lightlike, than $\delta s^2_{PQ}=0$.

Hence, $\delta s^2_{PQ} = 0$ if and only if relation $P\lightcone Q$ is fulfilled.

In the Minkowski space, where $g_{ik} = \eta_{ik}$, quantity $\delta s^2_{PQ}$ is the usual special-relativity interval:
\begin{equation}
\label{eq:2:delta-s-Minkowski}
\delta s^2_{PQ} = \eta_{ik} (r_Q^i - r_P^i) (r_Q^k - r_P^k) . \quad \text{(flat spacetime)}
\end{equation}
In a curved spacetime that is described by Eq.~(\ref{eq:2:g-series}), there must be corrections to Eq.~(\ref{eq:2:delta-s-Minkowski}). 
Let us define $\ve$-neighborhood of event $O$ as a set of such events $(t,x,y)$ that
\begin{equation}
\label{eq:2:neighborhood}
t^2 + x^2 + y^2 < \ve^2 .
\end{equation}
If both events $P$ and $Q$ belong to the $\ve$-neighborhood of event $O$, then, in the Riemann normal coordinates,~\cite{Brewin2009}
\begin{multline}
\label{eq:2:delta-s-via-R}
\delta s^2_{PQ} = \eta_{ik} (r_Q^i - r_P^i) (r_Q^k - r_P^k) \\
- \frac13 R_{iklm}(O)  \, r_P^i \, (r_Q^k - r_P^k)\, r_P^l (r_Q^m - r_P^m) \\
- \frac{1}{12} \left[\partial_n R_{iklm}(O)\right] r_P^i (r_Q^k - r_P^k) r_P^l (r_Q^m - r_P^m) (r_Q^n + r_P^n) \\
+ \mathcal O(\ve^6).
\end{multline}

\subsection{Two sets of events: ``unprimed'' and ``primed''}
\label{sec:12events}

Let is choose a small spatial scale $\ve$ (we will consider it as infinitesimal), and choose twelve events 1, 2, 3, 4, 5, 6, $a$, $b$, $c$, $d$, $e$, and $f$ in the $2\ve$-neighborhood of event $O$ as specified by Eqs.~(\ref{eq:well:135}) -- (\ref{eq:well:bdf}). 

If the spacetime were flat, then twenty three relations $a\lightcone1, \ldots, f\lightcone5$ listed in Definition~\ref{def:1} would be fulfilled. That is, the corresponding twenty three intervals $\delta s^2_{a1} \,, \ldots, \delta s^2_{f5}$ would be equal to zero. In a curved spacetime, this is only approximately true. If one substitutes the events' coordinates given by Eqs.~(\ref{eq:well:135}) -- (\ref{eq:well:bdf}) into Eq.~(\ref{eq:2:delta-s-via-R}), one can see that the first term in the right-hand side vanishes, but the other terms remain. Therefore, in the coordinate system chosen above, we have the following estimates for these intervals:
\begin{equation}
\label{eq:2:delta-s-estimate}
\delta s^2_{a1} = \mathcal O(\ve^4), \quad \ldots, \quad \delta s^2_{f5} = \mathcal O(\ve^4).
\end{equation}

The next goal is to find other twelve events $1'$, $2'$, $3'$, $4'$, $5'$, $6'$, $a'$, $b'$, $c'$, $d'$, $e'$, $f'$ that are close to events $1, \ldots, f$, but twenty three intervals between these new events $\delta s^2_{a'1'} \,, \ldots, \delta s^2_{f'5'}$ are exactly equal to zero. We can achieve this goal by the following six-step procedure.

Step 1. Events 1, 3, 4 and 5 remain unchanged, i.~e. we set events $1'=1$, $3'=3$, $4'=4$, and $5'=5$.

Step 2. We vary coordinates of events $a$ and $b$ a little, in order to set the intervals between them and events $1'$, $3'$ and $4'$ to zero. That is, we choose such events $a'$ and $b'$ in a close vicinity of events $a$ and $b$ correspondingly, that
\begin{subequations} \label{eq:2:delta-s-ab134}
\begin{gather}
\delta s^2_{a'1'} = \delta s^2_{a'3'} = \delta s^2_{a'4'} = 0, \label{eq:2:delta-s-a134} \\
\delta s^2_{b'1'} = \delta s^2_{b'3'} = \delta s^2_{b'4'} = 0. \label{eq:2:delta-s-b134} 
\end{gather}
\end{subequations}
This is possible because varying of three coordinates $t,x,y$ of event $a'$ is just enough to satisfy three conditions~(\ref{eq:2:delta-s-a134}), and the same is true for event $b'$. The exact meaning of wordings ``vary a little'', ``in a close vicinity'' will be discussed below.

Step 3. We choose events $e'$ and $f'$ in a close vicinity of events $e$ and $f$ correspondingly, such that
\begin{subequations} \label{eq:2:delta-s-ef345}
\begin{gather}
\delta s^2_{e'3'} = \delta s^2_{e'4'} = \delta s^2_{e'5'} = 0, \label{eq:2:delta-s-e345} \\
\delta s^2_{f'3'} = \delta s^2_{f'4'} = \delta s^2_{f'5'} = 0. \label{eq:2:delta-s-f345} 
\end{gather}
\end{subequations}
This step if fully analogous to Step 2.

Step 4. We choose such event $2'$ near event $2$, that conditions
\begin{equation}
\delta s^2_{a'2'} = \delta s^2_{b'2'} = 0 \label{eq:2:delta-s-2ab} 
\end{equation}
are satisfied. Since there are only two conditions, we can vary only two coordinates $x$ and $y$ of event $2'$, leaving its time-coordinate $t$ unchanged: $t_{2'} = t_2$.

Step 5. We choose events $c'$ and $d'$ in a close vicinity of events $c$ and $d$ correspondingly, such that
\begin{subequations} \label{eq:2:delta-s-cd125}
\begin{gather}
\delta s^2_{c'1'} = \delta s^2_{c'2'} = \delta s^2_{c'5'} = 0, \label{eq:2:delta-s-c125} \\
\delta s^2_{d'1'} = \delta s^2_{d'2'} = \delta s^2_{d'5'} = 0. \label{eq:2:delta-s-d125} 
\end{gather}
\end{subequations}

Step 6. We choose event $6'$ in a close vicinity of event $6$, to satisfy conditions
\begin{equation}
\delta s^2_{c'6'} = \delta s^2_{d'6'} = \delta s^2_{e'6'} = 0. \label{eq:2:delta-s-6cde} 
\end{equation}

Let us find out, how large are coordinate differences between ``primed'' events $1',\ldots,f'$ and their corresponding ``unprimed'' events $1,\ldots,f$. As an example, we consider the difference between events $a$ and $a'$. The difference between intervals $\delta s^2_{a'1'}$ and $\delta s^2_{a1'}$ can be estimated as
\begin{equation}
\label{eq:2:delta-s-difference-estim}
\delta s^2_{a'1'} - \delta s^2_{a1'} \approx \frac{\partial(\delta s^2_{a1'})}{\partial(r_a)^i} \, \left[ (r_{a'})^i - (r_a)^i \right] .
\end{equation}
The left-hand side of this equation is of order of $\mathcal O(\ve^4)$ due to Eqs.~(\ref{eq:2:delta-s-estimate}) and~(\ref{eq:2:delta-s-a134}). To estimate the derivative in the right-hand side, one can use Eq.~(\ref{eq:2:delta-s-via-R}):
\begin{multline}
\label{eq:2:delta-s-derivative-estim}
\frac{\partial(\delta s^2_{a1'})}{\partial(r_a)^i} \approx 
\frac {\partial}{\partial(r_a)^i} \, \eta_{jk} \left[ (r_a)^j - (r_{1'})^j \right] \left[ (r_a)^k - (r_{1'})^k \right] \\
= 2 \left[ (r_a)_i - (r_{1'})_i \right] = \mathcal O(\ve) .
\end{multline}
Therefore one can find from Eq.~(\ref{eq:2:delta-s-difference-estim}) that $(r_{a'})^i - (r_a)^i = \mathcal O(\ve^3)$. Similar considerations are valid for other events. Hence, we conclude that coordinates of ``primed'' events differ from that of corresponding ``unprimed'' events by no more than $\mathcal O(\ve^3)$.

Vanishing of intervals between the ``primed'' events in Eqs.~(\ref{eq:2:delta-s-ab134}) -- (\ref{eq:2:delta-s-6cde}) mean that twenty three relations
\begin{subequations} \label{eq:2:23-primed-relations}
\begin{gather}
  a'\lightcone1', \;\; a'\lightcone2', \;\; a'\lightcone3', \;\; a'\lightcone4', \\
  b'\lightcone1', \;\; b'\lightcone2', \;\; b'\lightcone3', \;\; b\lightcone4', \\
  c'\lightcone1', \;\; c'\lightcone2', \;\; c'\lightcone5', \;\; c\lightcone6', \\
  d'\lightcone1', \;\; d'\lightcone2', \;\; d'\lightcone5', \;\; d\lightcone6', \\
  e'\lightcone3', \;\; e'\lightcone4', \;\; e'\lightcone5', \;\; e\lightcone6', \\
  f'\lightcone3', \;\; f'\lightcone4', \;\; f'\lightcone5'
\end{gather}
\end{subequations}
are satisfied. The key question of this section is whether the twenty-fourth relation $f'\lightcone6'$ also holds. That is, whether the interval $\delta s^2_{f'6'}$ is equal to zero. We will show in the rest of this section that this interval is not equal to zero, that completes the proof of Theorem~\ref{th:Cotton}.

\subsection{Linear combination of intervals}
\label{sec:combination}

As we saw above, the intervals between the ``primed'' events $1',\ldots,f'$ differ from the corresponding intervals between the ``unprimed'' events $1,\ldots,f$ by $\mathcal O(\ve^4)$ at most. Here we will find such a linear combination of intervals, that its change during the transition from ``unprimed'' events to ``primed'' ones is as small as $\mathcal O(\ve^6)$. 

We define a linear combination $S$ as
\begin{equation}
\label{eq:2:S-def}
S = \sum_p \sum_q \gamma_{pq} \, \delta s^2_{pq} \, ,
\end{equation}
where symbol $p$ runs over events $a$, $b$, $c$, $d$, $e$, $f$; 
symbol $q$ runs over events 1, 2, 3, 4, 5, 6; and $\gamma_{pq}$ are some numbers. We search for such a set of coefficients $\gamma_{pq}$ that:

\noindent(i) only 24 coefficients $\gamma_{pq}$, for which the relations $p\lightcone q$ appear in Definition~\ref{def:1}, can differ from zero;

\noindent(ii) replacement of ``unprimed'' events $1, \ldots, f$ with their ``primed'' counterparts $1', \ldots, f'$ leaves the linear combination $S$ unchanged up to $\mathcal O(\ve^6)$, that is,
\begin{equation}
\label{eq:2:S-difference-1}
S' = S + \mathcal O(\ve^6) ,
\end{equation}
where
\begin{equation}
\label{eq:2:S-prime-def}
S' = \sum_p \sum_q \gamma_{pq} \, \delta s^2_{p'q'} \, .
\end{equation}

To achieve this goal, let us consider the difference between intervals $\delta s^2_{p'q'}$ and $\delta s^2_{pq}$. In the right-hand side of expression~(\ref{eq:2:delta-s-via-R}) for $\delta s^2_{pq}$, the first term has the order of magnitude $\mathcal O(\ve^2)$, the second one --- $\mathcal O(\ve^4)$, and the third one --- $\mathcal O(\ve^5)$. Coordinates $r_p^i$ and $r_q^i$ are of order of $\mathcal O(\ve)$, and their variations $(r_{p'}^i-r_p^i)$ and $(r_{q'}^i-r_q^i)$ are of order of $\mathcal O(\ve^3)$ as shown in Section~\ref{sec:12events}. Hence, the variation of the first term in Eq.~(\ref{eq:2:delta-s-via-R}) is $\mathcal O(\ve^4)$, of the second term --- $\mathcal O(\ve^6)$, and of the third term --- $\mathcal O(\ve^7)$. One can therefore keep the variation of only the first term in Eq.~(\ref{eq:2:delta-s-via-R}), and get
\begin{multline}
\label{eq:2:delta-s-difference-1}
\delta s^2_{p'q'} - \delta s^2_{pq} = \\
\eta_{ik} (r_{q'}^i - r_{p'}^i) (r_{q'}^k - r_{p'}^k) - \eta_{ik} (r_q^i - r_p^i) (r_q^k - r_p^k) + \mathcal O(\ve^6) .
\end{multline}
Then, writing down the coordinates $\vec{r}_{p'}$ and $\vec{r}_{q'}$ as
\begin{equation}
\label{eq:2:Delta-r}
\vec{r}_{p'} = \vec{r}_p + \overrightarrow{\Delta r}_p \, , \quad
\vec{r}_{q'} = \vec{r}_q + \overrightarrow{\Delta r}_q \, ,
\end{equation}
where coordinate differences $\overrightarrow{\Delta r}_p$ and $\overrightarrow{\Delta r}_q$ are of order of $\mathcal O(\ve^3)$, one can rewrite Eq.~(\ref{eq:2:delta-s-difference-1}) as
\begin{multline}
\label{eq:2:delta-s-difference-2}
\delta s^2_{p'q'} - \delta s^2_{pq} = \\
2 (\vec{r}_p - \vec{r}_q) \cdot \overrightarrow{\Delta r}_p - 2 (\vec{r}_p - \vec{r}_q) \cdot \overrightarrow{\Delta r}_q + \mathcal O(\ve^6) .
\end{multline}
Combination of equations~(\ref{eq:2:S-def}), (\ref{eq:2:S-prime-def}) and (\ref{eq:2:delta-s-difference-2}) gives rise to the following result:
\begin{multline}
\label{eq:2:S-difference-2}
S' - S = 
  2 \sum_p \sum_q \gamma_{pq} (\vec{r}_p - \vec{r}_q) \cdot \overrightarrow{\Delta r}_p \\
- 2 \sum_p \sum_q \gamma_{pq} (\vec{r}_p - \vec{r}_q) \cdot \overrightarrow{\Delta r}_q 
+ \mathcal O(\ve^6) .
\end{multline}
One can achieve the condition $S' = S + \mathcal O(\ve^6)$ by setting the coefficients at $\overrightarrow{\Delta r}_p$ and $\overrightarrow{\Delta r}_q$ to zero. This results in equations
\begin{equation}
\label{eq:2:equation-Delta-p}
\sum_q \gamma_{pq} (\vec{r}_p - \vec{r}_q) = 0 
\end{equation}
for $p = a, b, c, d, e, f$, and
\begin{equation}
\label{eq:2:equation-Delta-q}
\sum_p \gamma_{pq} (\vec{r}_p - \vec{r}_q) = 0 
\end{equation}
for $q = 1, 2, 3, 4, 5, 6$. More explicitly, these equations read:
\begin{gather}
  \gamma_{a1} (\vec{r}_a \!\!-\!\! \vec{r}_1) + \gamma_{a2} (\vec{r}_a \!\!-\!\! \vec{r}_2) + \gamma_{a3} (\vec{r}_a \!\!-\!\! \vec{r}_3) + \gamma_{a4} (\vec{r}_a \!\!-\!\! \vec{r}_4) = 0, \nonumber\\
  \gamma_{b1} (\vec{r}_b \!\!-\!\! \vec{r}_1) + \gamma_{b2} (\vec{r}_b \!\!-\!\! \vec{r}_2) + \gamma_{b3} (\vec{r}_b \!\!-\!\! \vec{r}_3) + \gamma_{b4} (\vec{r}_b \!\!-\!\! \vec{r}_4) = 0, \nonumber\\
  \gamma_{c1} (\vec{r}_c \!\!-\!\! \vec{r}_1) + \gamma_{c2} (\vec{r}_c \!\!-\!\! \vec{r}_2) + \gamma_{c5} (\vec{r}_c \!\!-\!\! \vec{r}_5) + \gamma_{c6} (\vec{r}_c \!\!-\!\! \vec{r}_6) = 0, \nonumber\\
  \gamma_{d1} (\vec{r}_d \!\!-\!\! \vec{r}_1) + \gamma_{d2} (\vec{r}_d \!\!-\!\! \vec{r}_2) + \gamma_{d5} (\vec{r}_d \!\!-\!\! \vec{r}_5) + \gamma_{d6} (\vec{r}_d \!\!-\!\! \vec{r}_6) = 0, \nonumber\\
  \gamma_{e3} (\vec{r}_e \!\!-\!\! \vec{r}_3) + \gamma_{e4} (\vec{r}_e \!\!-\!\! \vec{r}_4) + \gamma_{e5} (\vec{r}_e \!\!-\!\! \vec{r}_5) + \gamma_{e6} (\vec{r}_e \!\!-\!\! \vec{r}_6) = 0, \nonumber\\
  \gamma_{f3} (\vec{r}_f \!\!-\!\! \vec{r}_3) + \gamma_{f4} (\vec{r}_f \!\!-\!\! \vec{r}_4) + \gamma_{f5} (\vec{r}_f \!\!-\!\! \vec{r}_5) + \gamma_{f6} (\vec{r}_f \!\!-\!\! \vec{r}_6) = 0, \nonumber\\
  \gamma_{a1} (\vec{r}_a \!\!-\!\! \vec{r}_1) + \gamma_{b1} (\vec{r}_b \!\!-\!\! \vec{r}_1) + \gamma_{c1} (\vec{r}_c \!\!-\!\! \vec{r}_1) + \gamma_{d1} (\vec{r}_d \!\!-\!\! \vec{r}_1) = 0, \nonumber\\
  \gamma_{a2} (\vec{r}_a \!\!-\!\! \vec{r}_2) + \gamma_{b2} (\vec{r}_b \!\!-\!\! \vec{r}_2) + \gamma_{c2} (\vec{r}_c \!\!-\!\! \vec{r}_2) + \gamma_{d2} (\vec{r}_d \!\!-\!\! \vec{r}_2) = 0, \nonumber\\
  \gamma_{a3} (\vec{r}_a \!\!-\!\! \vec{r}_3) + \gamma_{b3} (\vec{r}_b \!\!-\!\! \vec{r}_3) + \gamma_{e3} (\vec{r}_e \!\!-\!\! \vec{r}_3) + \gamma_{f3} (\vec{r}_f \!\!-\!\! \vec{r}_3) = 0, \nonumber\\
  \gamma_{a4} (\vec{r}_a \!\!-\!\! \vec{r}_4) + \gamma_{b4} (\vec{r}_b \!\!-\!\! \vec{r}_4) + \gamma_{e4} (\vec{r}_e \!\!-\!\! \vec{r}_4) + \gamma_{f4} (\vec{r}_f \!\!-\!\! \vec{r}_4) = 0, \nonumber\\
  \gamma_{c5} (\vec{r}_c \!\!-\!\! \vec{r}_5) + \gamma_{d5} (\vec{r}_d \!\!-\!\! \vec{r}_5) + \gamma_{e5} (\vec{r}_e \!\!-\!\! \vec{r}_5) + \gamma_{f5} (\vec{r}_f \!\!-\!\! \vec{r}_5) = 0, \nonumber\\
  \gamma_{c6} (\vec{r}_c \!\!-\!\! \vec{r}_6) + \gamma_{d6} (\vec{r}_d \!\!-\!\! \vec{r}_6) + \gamma_{e6} (\vec{r}_e \!\!-\!\! \vec{r}_6) + \gamma_{f6} (\vec{r}_f \!\!-\!\! \vec{r}_6) = 0. \label{eq:2:12equations}
\end{gather}
This is a set of 36 linear equations (12 ones per each of coordinates $t, x, y$) with 24 unknowns $\gamma_{a1}\,, \ldots, \gamma_{f6}$. Event coordinates $\vec{r}_1\,, \ldots, \vec{r}_f$ are to be taken from Eqs.~(\ref{eq:well:135}) -- (\ref{eq:well:bdf}).

One can rewrite this system of equations in a matrix form $\mathcal M \, \Gamma = 0$ with a $36 \times 24$ matrix $\mathcal M$ and a column vector $\Gamma$ of 24 unknown values $\gamma_{a1}\,, \ldots, \gamma_{f6}$. Then, it can be checked numerically (e.~g. with the aid of MATLAB) that the rank of matrix $\mathcal M$ is equal to 23. This means that there are only 23 linearly-independent equations in the set~(\ref{eq:2:12equations}). One can therefore give an arbitrary value to one of 24 unknown variables. For definiteness, we set $\gamma_{f6} = 1$. Then, the values of other 23 coefficients $\gamma_{a1}\,, \ldots, \gamma_{f5}$ can be found from equations~(\ref{eq:2:12equations}). The obtained values are listed in Table~\ref{table:gamma}.
\begin{table}
\centering
\begin{tabular}{ c|rrrrrr| } 
  & 1 & 2 & 3 & 4 & 5 & 6 \\ 
 \hline
 $a$ & 2 &-2 &-2 & 2 & 0 & 0 \\ 
 $b$ & 1 &-1 &-1 & 1 & 0 & 0 \\ 
 $c$ &-2 & 2 & 0 & 0 & 2 &-2 \\ 
 $d$ &-1 & 1 & 0 & 0 & 1 &-1 \\ 
 $e$ & 0 & 0 & 2 &-2 &-2 & 2 \\ 
 $f$ & 0 & 0 & 1 &-1 &-1 & 1 \\ 
 \hline
\end{tabular}
\caption{Coefficients $\gamma_{pq}$ for $p = a, b, c, d, e, f$ and $q = 1, 2, 3, 4, 5, 6$.}
\label{table:gamma}
\end{table}

Thus, we have found such a set of coefficients $\gamma_{pq}$ (see Table~\ref{table:gamma}) that linear combination of intervals $S$ remains unchanged up to $\mathcal O(\ve^6)$ under replacement of the ``unprimed'' events with the ``primed'' ones. That is, condition~(\ref{eq:2:S-difference-1}) holds.

\subsection{Calculation of the interval between events $f'$ and $6'$}
\label{sec:calculation-ds-f6}

In Section~\ref{sec:combination}, we have found such a set of coefficients $\gamma_{pq}$ (see Table~\ref{table:gamma}), that enables equality $S' = S + \mathcal O(\ve^6)$. Let us now simplify the expression~(\ref{eq:2:S-prime-def}) for quantity $S'$. The sum in Eq.~(\ref{eq:2:S-prime-def}) consists of only one nonzero term $\gamma_{f6}\,\delta s^2_{f'6'}$. Indeed, for all but one pairs $(p,q)$ in this sum, either the interval $\delta s^2_{p'q'}$ vanishes due to Eqs.~(\ref{eq:2:delta-s-ab134}) -- (\ref{eq:2:delta-s-6cde}), or the coefficient $\gamma_{pq}$ vanishes according to Table~\ref{table:gamma}. The only exception is the pair $(f,6)$. Taking $\gamma_{f6}=1$ from Table~\ref{table:gamma}, one obtains
\begin{equation}
\label{eq:2:S-prime-1}
S' = \delta s^2_{f'6'} \, .
\end{equation}

Combination of equations~(\ref{eq:2:S-def}), (\ref{eq:2:S-difference-1}) and~(\ref{eq:2:S-prime-1}) allows us to express the ``primed'' interval $\delta s^2_{f'6'}$ through ``unprimed'' intervals $\delta s^2_{a1}\,, \ldots \delta s^2_{f6}$:
\begin{equation}
\label{eq:2:delta-s-f6-1}
\delta s^2_{f'6'} = \sum_p \sum_q \gamma_{pq} \, \delta s^2_{pq} + \mathcal O(\ve^6) .
\end{equation}

Then let us evaluate the ``unprimed'' intervals by substituting the coordinates of ``unprimed'' events from Eqs.~(\ref{eq:well:135}) -- (\ref{eq:well:bdf}) into Eq.~(\ref{eq:2:delta-s-via-R}). To this end, it is convenient to rewrite Eq.~(\ref{eq:2:delta-s-via-R}) as
\begin{equation}
\label{eq:2:delta-s-via-R-123}
\delta s^2_{pq} = (\delta s^2_{pq})^{(1)} + (\delta s^2_{pq})^{(2)} + (\delta s^2_{pq})^{(3)} + \mathcal O(\ve^6),
\end{equation}
where
\begin{equation}
\label{eq:2:delta-s-via-R-1}
(\delta s^2_{pq})^{(1)} = \eta_{ik} (r_q^i - r_p^i) (r_q^k - r_p^k) , 
\end{equation}
\begin{equation}
\label{eq:2:delta-s-via-R-2}
(\delta s^2_{pq})^{(2)} = - \frac13 R_{iklm}(O)  \, r_p^i \, (r_q^k - r_p^k)\, r_p^l (r_q^m - r_p^m) , 
\end{equation}
\begin{multline}
\label{eq:2:delta-s-via-R-3}
(\delta s^2_{pq})^{(3)} = \\
- \frac{1}{12} \left[\partial_n R_{iklm}(O)\right] r_p^i (r_q^k - r_p^k) r_p^l (r_q^m - r_p^m) (r_q^n + r_p^n) .
\end{multline}
The value $(\delta s^2_{pq})^{(1)}$ is just an interval between events $p$ and $q$ in a flat Minkowski spacetime. With coordinates of the ``unprimed'' events given by Eqs.~(\ref{eq:well:135}) -- (\ref{eq:well:bdf}), and coefficients $\gamma_{pq}$ from Table~\ref{table:gamma}, one can easily see that for each pair $(p,q)$ either $\gamma_{pq} = 0$ or $(\delta s^2_{pq})^{(1)} = 0$. Hence,
\begin{equation}
\label{eq:2:S-1}
\sum_p \sum_q \gamma_{pq} \, (\delta s^2_{pq})^{(1)} = 0 .
\end{equation}
The interval $\delta s^2_{f'6'}$ is therefore a sum of linear combinations of terms $(\delta s^2_{pq})^{(2)}$ and $(\delta s^2_{pq})^{(2)}$. These combinations can be represented as
\begin{equation}
\label{eq:2:S-2}
\sum_p \sum_q \gamma_{pq} \, (\delta s^2_{pq})^{(2)} = \ve^4 M^{iklm} R_{iklm}(O) 
\end{equation}
and
\begin{equation}
\label{eq:2:S-3}
\sum_p \sum_q \gamma_{pq} \, (\delta s^2_{pq})^{(3)} = \ve^5 N^{iklmn} \partial_n R_{iklm}(O) ,
\end{equation}
where tensors $M^{iklm}$ and $N^{iklmn}$ are combinations of coordinates of ``unprimed'' events, as they appear in Eqs.~(\ref{eq:2:delta-s-via-R-2}) and~(\ref{eq:2:delta-s-via-R-3}):
\begin{equation}
\label{eq:2:M-iklm}
M^{iklm} = - \frac{1}{3\ve^4} \sum_p \sum_q \gamma_{pq} \,  r_p^i \, (r_q^k - r_p^k)\, r_p^l (r_q^m - r_p^m) , 
\end{equation}
\begin{multline}
\label{eq:2:N-iklmn}
N^{iklmn} = \\
- \frac{1}{12\ve^5} \sum_p \sum_q \gamma_{pq} \,  r_p^i (r_q^k - r_p^k) r_p^l (r_q^m - r_p^m) (r_q^n + r_p^n) .
\end{multline}

Gathering Eqs.~(\ref{eq:2:delta-s-f6-1}), (\ref{eq:2:delta-s-via-R-123}), (\ref{eq:2:S-1}) -- (\ref{eq:2:S-3}) together, we relate the interval $\delta s^2_{f'6'}$ to the Riemann tensor $R_{iklm}$ and its derivatives:
\begin{multline}
\label{eq:2:delta-s-f6-2}
\delta s^2_{f'6'} = \sum_p \sum_q \gamma_{pq} \, (\delta s^2_{pq})^{(2)} + \sum_p \sum_q \gamma_{pq} \, (\delta s^2_{pq})^{(3)} + \mathcal O(\ve^6) \\
= \ve^4 M^{iklm} R_{iklm}(O) + \ve^5 N^{iklmn} \partial_n R_{iklm}(O) + \mathcal O(\ve^6) .
\end{multline}

Tensors $M^{iklm}$ and $N^{iklmn}$ can be evaluated by substituting of coordinates of ``unprimed'' events from Eqs.~(\ref{eq:well:135}) -- (\ref{eq:well:bdf}) into Eqs.~(\ref{eq:2:M-iklm}) and~(\ref{eq:2:N-iklmn}). This arithmetic job has been done with the aid of a computer. As a result, we have found that almost all components of tensors $M^{iklm}$ and $N^{iklmn}$ are equal to zero. Non-vanishing components of tensor $M^{iklm}$ are
\begin{gather*}
M^{xtxx} = M^{xxxt} = 1 , \\
M^{xtyy} \!=\! M^{yyxt} \!=\! M^{xyyt} \!=\! M^{ytxy} \!=\! M^{ytyx} \!=\! M^{yxyt} \!=\! -1 ,
\end{gather*}
and non-vanishing components of tensor $N^{iklmn}$ are
\begin{gather*}
N^{xtxxy} = N^{xxxty} = -N^{ytyyx} = -N^{yyytx} = \frac{3}{8\sqrt3} \,, \\
N^{ytyxy} = N^{yxyty} = -N^{xtxyx} = -N^{xyxtx} = \frac{5}{8\sqrt3} \,, \\
N^{xxxyt} = N^{xyxxt} = N^{yyxyt} = N^{xyyyt} = \\
= N^{xxytx} = N^{ytxxx} = N^{xtyxx} = N^{yxxtx} = \frac{1}{8\sqrt3} \,, \\
N^{yyyxt} = N^{yxyyt} = N^{xxyxt} = N^{yxxxt} = \\
= N^{yyxty} = N^{xtyyy} = N^{ytxyy} = N^{xyyty} = -\frac{1}{8\sqrt3} \,.
\end{gather*}
Equation~(\ref{eq:2:delta-s-f6-2}) thus acquires the following form:
\begin{widetext}
\begin{multline}\label{eq:2:delta-s-f6-3}
\delta s^2_{f'6'} = 
\ve^4 \big[ R_{xtxx}(O) + R_{xxxt}(O) - R_{xtyy}(O) - R_{yyxt}(O) 
- R_{xyyt}(O) - R_{ytxy}(O) - R_{ytyx}(O) - R_{yxyt}(O) \big] \\
+ \frac{\ve^5}{8\sqrt3} \big[ 3 \partial_y R_{xtxx}(O) + 3 \partial_y R_{xxxt}(O) 
- 3 \partial_x R_{ytyy}(O) - 3 \partial_x R_{yyyt}(O) + 5 \partial_y R_{ytyx}(O) 
+ 5 \partial_y R_{yxyt}(O) - 5 \partial_x R_{xtxy}(O) - 5 \partial_x R_{xyxt}(O) \\
+ \partial_t R_{xxxy}(O) + \partial_t R_{xyxx}(O) + \partial_t R_{yyxy}(O) + \partial_t R_{xyyy}(O) 
+ \partial_x R_{xxyt}(O) + \partial_x R_{ytxx}(O) + \partial_x R_{xtyx}(O) + \partial_x R_{yxxt}(O) \\
- \partial_t R_{yyyx}(O) - \partial_t R_{yxyy}(O) - \partial_t R_{xxyx}(O) - \partial_t R_{yxxx}(O) 
- \partial_y R_{yyxt}(O) - \partial_y R_{xtyy}(O) - \partial_y R_{ytxy}(O) - \partial_y R_{xyyt}(O) \big] 
+ \mathcal O(\ve^6) .
\end{multline}
The majority of terms in Eq.~(\ref{eq:2:delta-s-f6-3}) vanish due to skew symmetry of the Riemann tensor. This symmetry dictates that tensor components with the same 1st and 2nd indices are equal to zero: $R_{iikl}=0$, as well as components with the same 3rd and 4th indices: $R_{ikll}=0$. Let us remove such vanishing components from Eq.~(\ref{eq:2:delta-s-f6-3}):
\begin{multline}\label{eq:2:delta-s-f6-4}
\delta s^2_{f'6'} = 
\ve^4 \big[ - R_{xyyt}(O) - R_{ytxy}(O) - R_{ytyx}(O) - R_{yxyt}(O) \big] 
+ \frac{\ve^5}{8\sqrt3} \big[ 5 \partial_y R_{ytyx}(O) + 5 \partial_y R_{yxyt}(O) \\
- 5 \partial_x R_{xtxy}(O) - 5 \partial_x R_{xyxt}(O) 
+ \partial_x R_{xtyx}(O) + \partial_x R_{yxxt}(O) - \partial_y R_{ytxy}(O) - \partial_y R_{xyyt}(O) \big] 
+ \mathcal O(\ve^6) .
\end{multline}
\end{widetext}
One can further simplify this equation by employing the symmetry properties of the Riemann tensor,
\begin{equation}\label{eq:2:Riemann-symmetry}
R_{iklm} = R_{lmik} \, , \quad
R_{iklm} = -R_{kilm} \, , \quad
R_{iklm} = -R_{ikml} \, , 
\end{equation}
and obtain:
\begin{equation}\label{eq:2:delta-s-f6-5}
\delta s^2_{f'6'} = 
\frac{\sqrt3}{2} \, \ve^5 \big[ \partial_y R_{yxyt}(O) - \partial_x R_{xyxt}(O) \big] 
+ \mathcal O(\ve^6) .
\end{equation}
Here, the expression within the square brackets is equal to the right-hand side of Eq.~(\ref{eq:2:Ytt-at-O}). Let us replace it with the left-hand side:
\begin{equation}\label{eq:2:delta-s-f6-6}
\delta s^2_{f'6'} = 
\frac{\sqrt3}{2} \, \ve^5 \, Y_{tt}(O) + \mathcal O(\ve^6) .
\end{equation}
Hence, we have reduced the interval $\delta s^2_{f'6'}$ to the $tt$-component of the Cotton-York tensor.

\subsection{Finalization of the proof}
\label{sec:theorem2-final}

Let us recall the content of the above parts of this section. We consider a 3-dimensional spacetime that has a non-zero Cotton tensor at some point $O$. We are going to prove that this spacetime is not well-stitched. A natural way to do it is to provide a counterexample---a set of twelve events that violates well-stitchedness. Below we argue that ``primed'' events $1', \ldots, f'$ constitute such a counterexample.

In Section~\ref{sec:york}, we have shown that it is possible to find a reference frame, in which the component $Y_{tt}$ of the Cotton-York tensor is non-zero at event $O$. In Section~\ref{sec:frame}, we have chosen Riemann normal coordinates that obey the condition $Y_{tt}(O) \neq 0$. In Section~\ref{sec:intervals}, we have considered the ``interval'' $\delta s^2_{PQ}$ between events $P$ and $Q$ in a curved spacetime. This interval is equal to zero if and only if the relation $P \lightcone Q$ is fulfilled. In Section~\ref{sec:12events}, we have introduced two sets of events: ``unprimed'' events $1, \ldots, f$ that are depicted in Fig.~\ref{fig:simple-set}, and ``primed'' events $1', \ldots, f'$ that are located near their ``unprimed'' counterparts. The ``primed'' events are chosen in such a way that 23 intervals between them $\delta s^2_{a'1'} \,, \ldots, \delta s^2_{f'5'}$ vanish, and hence 23 relations $a' \lightcone 1', \ldots, f' \lightcone 5'$ are fulfilled. In Section~\ref{sec:combination}, we have found a linear combination of intervals $S$, which remains unchanged up to $\mathcal O(\ve^6)$ when ``unprimed'' events are replaced with ``primed'' ones. (Here parameter $\ve$ is a characteristic size of the set of events.) Using this result, in Section~\ref{sec:calculation-ds-f6} we have reduced the interval $\delta s^2_{f'6'}$ to the $tt$-component of the Cotton-York tensor, as stated by Eq.~(\ref{eq:2:delta-s-f6-6}).

Then we go further and make a final step of the proof. To this end, it is convenient to rewrite Eq.~(\ref{eq:2:delta-s-f6-6}) as follows:
\begin{equation}\label{eq:2:delta-s-f6-7}
\delta s^2_{f'6'} = 
\frac{\sqrt3}{2} \, \ve^5 \big[ Y_{tt}(O) + \mathcal O(\ve) \big] .
\end{equation}
Choosing small enough size $\ve$, one can ensure that the right-hand side of this equation is different from zero. Indeed, $Y_{tt}(O) \neq 0$ according to our choice of coordinates in Section~\ref{sec:frame}, and the remainder $\mathcal O(\ve)$ is smaller than $Y_{tt}(O)$ for a small enough value of $\ve$. Hence, there exists such a size $\ve$ that
\begin{equation}\label{eq:2:delta-s-f6-8}
\delta s^2_{f'6'} \neq 0 .
\end{equation}

Now it is evident that, for a sufficiently small parameter $\ve$, events $1', 2', 3', 4', 5', 6', a', b', c', d', e'$ and $f'$ provide a desired counterexample that proves Theorem~\ref{th:Cotton}. Let us argue by contradiction. Suppose that the spacetime is well-stitched. Then, by virtue of Definition~\ref{def:1}, it would follow from relations~(\ref{eq:2:23-primed-relations}) that relation $f' \lightcone 6'$ is fulfilled. But, as it follows from Eq.~(\ref{eq:2:delta-s-f6-8}), relation $f' \lightcone 6'$ is violated. We therefore have come to a contradiction. Hence, the spacetime, in which the Cotton tensor does not vanish at some point $O$, is not well-stitched. Theorem~\ref{th:Cotton} is thus proved.

\


\section{Conclusions}
\label{sec:conclusions}

In this paper we have spread the findings of previous work~\cite{Nenashev2023} to the three-dimensional (3D) pseudo-Riemannian spacetime, which includes 2 spatial + 1 temporal coordinates. 
We have found that the flat Minkowski 3D spacetime, as well as any conformally-flat 3D spacetime, possesses a structure of 24 causal relations between 12 events. This structure is described in Definition~\ref{def:1} as a feature of a ``well-stitched spacetime''. 
We have proved (see Theorems~\ref{th:Minkowski} and~\ref{th:Cotton} and their proofs in Sections~\ref{sec:proof1} and~\ref{sec:proof2}) that a 3D spacetime is ``well-stitched'' if and only if it is conformally flat.

We emphasize that the definition of a ``well-stitched'' spacetime does not contain any metrical information: no lengths, no times, only the causal relations of type ${\mathrm A} \lightcone {\mathrm B}$. Hence, the concept of a ``well-stitched'' spacetime does not belong to the metric geometry, but rather to geometry of incidence. We therefore have ``translated'' an important concept of a conformally-flat spacetime from the ``metric'' language of Riemannian geometry to the ``non-metric'' language of the geometry of incidence.

The results of this paper provide a tool for detecting the curvature of the 3D spacetime on the basis of causal relations only, without any measurement instruments like rulers and clocks, provided that the spacetime is not conformally flat. This tool consists in finding such twelve events 1, 2, 3, 4, 5, 6, $a$, $b$, $c$, $d$, $e$, $f$ that 23 relations $a\lightcone1, \ldots, f\lightcone5$ listed in Definition~\ref{def:1} are fulfilled. If the twenty-fourth relation $f\lightcone6$ appears to be \emph{not} fulfilled, then according to Definition~\ref{def:1} the spacetime is not well-stitched, and due to Theorem~\ref{th:Minkowski} it must be curved. If, on the contrary, the relation $f\lightcone6$ is always fulfilled in this setting, then the spacetime is well-stitched, and consequently it is conformally flat by the virtue of Corollary~\ref{cor:Cotton}.


\bibliography{connection}

\end{document}